\documentclass[aps,prl,twocolumn,floatfix]{revtex4}
\usepackage{amssymb,amsmath,amsfonts}
\usepackage{graphicx,graphics}
\usepackage{epsfig}
\usepackage{blkarray}
\usepackage{epstopdf}
\usepackage[FIGBOTCAP]{subfigure}
\usepackage{amsfonts}
\usepackage{bm,bbm}
\usepackage{amssymb,amsmath,amsfonts}
\usepackage{mathrsfs}
\usepackage{calc}
\usepackage{epsfig}
\usepackage{float}
\usepackage{color}
\usepackage{epstopdf}
\usepackage{bm,amssymb}
\usepackage{graphicx,color}
\usepackage{comment}
\usepackage{soul}
\usepackage{amsmath}
\newcommand{\be}{\begin{equation}}
\newcommand{\ee}{\end{equation}}

\newcommand{\beqar}{\begin{eqnarray}}
\newcommand{\eeqar}{\end{eqnarray}}
\newcommand{\bcen}{\begin{center}}
\newcommand{\ecen}{\end{center}}

\DeclareUnicodeCharacter{2212}{-}
\begin{document}

\title{Photon Statistics from Non-Hermitian Floquet Theory: High Harmonic Generation and Above-Threshold Ionization Spectra Detected via IR Detectors}
\author{Nimrod Moiseyev$^{1,2,3}$ \footnote{\tt nimrod@technion.ac.il https://nhqm.net.technion.ac.il/} }

\affiliation{Schulich Faculty of Chemistry$^1$, Faculty of Physics$^2$ and Solid State Institute$^3$, Technion-Israel Institute of Technology, Haifa 32000, Israel}

\begin{abstract}
Although it seems that obtaining quantum properties of light from classical calculations is a self-contradictory claim, it is shown here that a unified mechanism governs the three distinct measurements of high harmonic generation spectra (HGS), above-threshold ionization (ATI), and IR photon number distribution, none of which require the quantization of the electromagnetic field.  
 Here, the conditions that enable the calculations of HGS and ATI spectra for atoms interacting with high-intensity laser fields from photon statistics are first derived. Through the non-Hermitian theoretical simulation, the regimes where there is correspondence between the HHG and ATI spectra and annihilated pump photons (with post-selection) are identified.  Consequently, the HGS and ATI spectra, as detected by XUV detectors, can be obtained by monitoring the fluctuations of the infrared absorbed photons.  
\end{abstract}

\maketitle
\section{Introduction and motivation}

The groundbreaking research by L'Huillier\cite{HHG-LeHuillier} presented experimental results on high harmonic generation (HHG) using a ultra high intensity infrared laser, leading to the development of Atto-science.
The high order harmonics are VUV radiation with the frequency $\omega_{XUV}=\omega_{IR}N_{IR}$ which are emitted due to the absorption of $N_{IR}$ infrared photons.
Recently in  Refs.\citenum{photon_statistics2017,gonoskov2016quantum} it was demonstrated that the high harmonics spectra emitted by argon and xenon atoms is imprinted in the remaining intensity distribution of infrared radiation after taking into account correlations with the XUV photon numbers.
 Their post selective analysis was based on a requirement that the total energy of light (XUV and IR) is conserved, i.e., the HGS process is parametric and treated the electromagnetic field quantum mechanically. The experiment claimed to demonstrate the quantum-optical nature of the high-order harmonic generation.
 Here we provide a Floquet perspective of their measurements, that does not require quantization of light. We interpret the experimental scheme as a post-selection on a single resonance (non-hermitian) Floquet solution. \textit{This condition is in agreement with the calculations of the  HGS and photon statistics from a single Floquet solution}. This can be done only in NHQM when the resonances are in the discrete part of the spectrum of the Floquet operator\cite{NHQM-BOOK}. The Floquet operator consists the atomic Hamiltonian that interacts with the classical electromagnetic field. Below, using a \textit{single} complex resonance Floquet solution, we will prove  that HGS can be deduced from the photon statistics of the absorbed number N of IR Floquet "photons", without quantization of light due to the discrete time transformation symmetry. As we have shown already  when the number of absorbed Floquet "photons" is significantly smaller than the total number of photons emitted by the strong laser, one would anticipate that similar results will be obtained from quantum electrodynamics (QED) calculations, where the light is treated quantum mechanically \cite{moiseyev-Even-Tzur2023}.
{ Notice that while in Ref. \citenum{gonoskov2016quantum} a semi-classical approach (known as three-step models) was used for the description of the HHG process, here the theory is based on the exact solution of the time-periodic Hamiltonian. For the sake of simplicity in the calculations, a model Hamiltonian for Xenon in strong laser fields has been used. In Ref. \citenum{gonoskov2016quantum}, Fock states were used as a basis set. We have already proven \cite{moiseyev-Even-Tzur2023} that when the bandwidth of Fourier channels contributing to the resonance Floquet solution is narrow, the occupations quantum Fock states can be approximated by the resonance Floquet solutions obtained through complex scaling transformation. Using the non-Hermitian approach, a rigorous proof for the equality between HHG and photon statistics spectra has been derived.}
\section{Study of the correspondence between HGS and photon statistics from a single resonance Floquet solution}

Since this work focuses on the interactions of atoms with CW laser fields, described by time-periodic potentials within the framework of the classical electromagnetic field approach, and on photon statistics to explore the quantum nature of light, let me start by mentioning works that, a priory, seem difficult to bridge between the two and obtain quantum light information using the classical light approach.

\subsubsection{Photon Statistics}
Photon statistics are playing a central role in different fields of research.  See for example below several of them:

$\bullet$ Ref.\citenum{PHOTON-STATISTICS-2023-LEWENSTEIN} on nonlinear optics using intense optical Schr\"odinger cat states.\\
$\bullet$ Ref.\citenum{PHPTON-STATISTICS-2014-HGS} on
Quantum-optical nature of the recollision process in high-order-harmonic generation.\\
$\bullet$ Ref.\citenum{PHOTON-STATISTICS-2023-compton} on
Compton scattering driven by intense quantum light.\\
$\bullet$ Ref.\citenum{PHOTON-STATISTICS-1983-negative-mandel-parameter} on First observation of sub-Poissonian (negative Mandel parameter) photon statistics.\\
 $\bullet$ Ref.\citenum{PHOTON-STATISTICS-1990-observation-negative-mandel-parameter} on the observation of sub-Poissonian photon statistics (negative Mandel parameter) in a micromaser.\\
  $\bullet$ Ref.\citenum{PHOTON-STATISTICS-2019-accurate-detection} on
  an accurate detection of arbitrary photon statistics.\\
  $\bullet$ Ref.\citenum{PHOTON-STATISTICS-FORCE-2023-MATAN-in-ultrafast-electron-dynamics} on Photon-statistics force in ultrafast electron dynamics.\\
$\bullet$ Ref.\citenum{HGS-from-quantumlight-2023-KAMINER} on
  high-harmonic generation driven by quantum light.\\
  $\bullet$ Ref.\citenum{QUANTUM-LIGHT-2024-MATAN-motion-charged-particls} on motion of charged particles in bright squeezed vacuum.

\subsubsection{ Periodic Floquet theory}

The Hamiltonian that describes the interaction of an atom with strong CW IR laser fields is time-periodic. The Floquet solutions of the time-dependent Schr\"odinger equation are given by 
$|\psi_E\rangle=e^{-iEt/hbar}\sum_{n=-\infty}^{+\infty} e^{i\omega_{IR} nt} |\varphi_n\rangle$, 
where $\omega_{IR}$ is the frequency of the cw laser field. The Floquet channel $n=0$ represents the atom in a strong laser field consisting of stands for the atom in strong laser field that consists $N_p$ photons. $N_p$ can be hundred's thousands of photons.  Under specific conditions presented in Ref. \citenum{moiseyev-Even-Tzur2023}, $n>0$  represents the number of absorbed IR photons that emit the XUV radiation with photon energies $\hbar\Omega_n$, where $\Omega_n=n\omega_{IR}$. The Floquet theory is applied to a wide range of physical phenomena, and below we mention only several references that provide the mathematical background of Floquet theory in addition to its use for calculating HGS.

$\bullet$ Ref.\citenum{faisal1997floquet} on Floquet-Bloch theory of high-harmonic generation in periodic structures.\\

$\bullet$  Ref.\citenum{casas2001floquet} on Floquet theory: exponential perturbative treatment.

\subsection{Theory}
The post-selection made in Ref.\citenum{photon_statistics2017} on some diagonal in their diagram  seems to select the points where the total energy of XUV and IR light is conserved.  This fits to our calculations of the HGS form a singe resonance Floquet solution that is associated with the energy of a mixed state of the atom and light\cite{NM-FW-PRL}. Notice that only when the resonance Floquet solutions are discrete
and therefore only then the HGS and ATI can be calculated from a single Floquet solution.
 A single resonance Floquet state that is dominated by the ground, field-free electronic state of an atom allows for the derivation of both high-order harmonics, ATI,  and photon statistics. This is achieved through the probability amplitude to transfer $N$ infrared Floquet "photons" from the n-Floquet channel to the (n+N)-Floquet channel of the resonance Floquet state. See Ref.\citenum{alon1998selection} and   Ref.\citenum{NM-Weinhold-PRL} for the first agreement between ab-inition calculations of the HGS and experimental measuremnt of the HGS. In Ref.\citenum{NM-Lien2003} it is shown that this approach which is  based on  non-hermitian calculations of the resonance Floquet solutions provides  \textit{the same} HGS as calculated by the standard (Hermitian) quantum mechanics calculations.
 Let us focus now on the introducing the probability amplitude to transfer $N$ infrared photons from the n-Floquet channel to the (n+N)-Floquet channel of the resonance Floquet state which is the building block in the calculations of HGS, ATI and photon statistics. Notice that this probability amplitude is not equal to zero iff N gets odd values only.
\subsubsection{The probability amplitude to transfer $N$ infrared photons from one Floquet channel to another}
\begin{figure}[h!]
     \centering
     \includegraphics[width=\columnwidth] {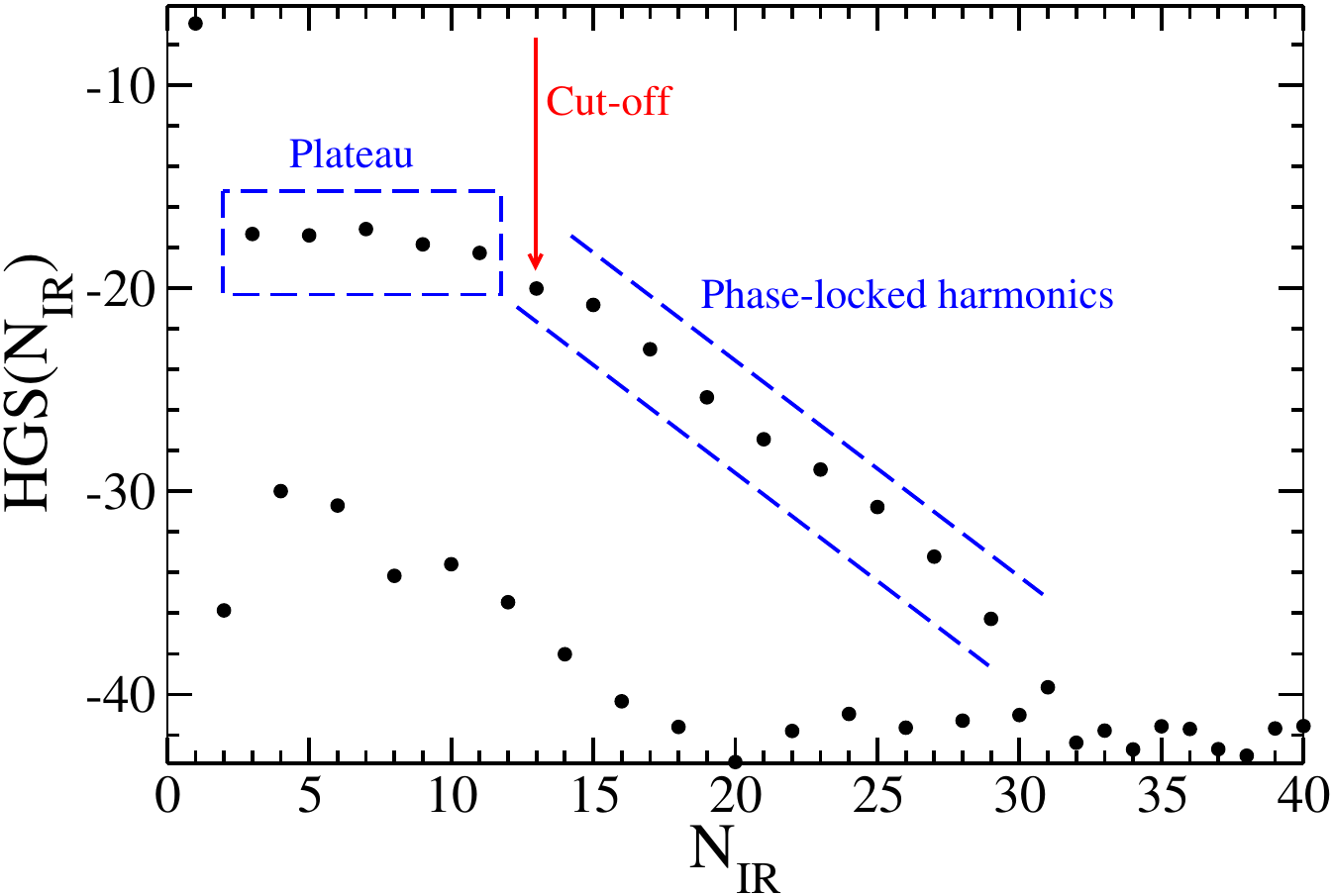}
      \caption{The HGS as a function of the number of absorbed IR photons, on log scale. 
   Notice that $N_{IR}=\omega_{XUV}/\omega_{IR}$ and therefore as discussed in the text this spectra can be measured from infrared photon statistics.  }
    \label{HGS-Xe}
    \end{figure}

The Floquet Hamiltonian in the length guage representation is given by $$\hat H_f\equiv -i\hbar \partial_t +\hat H_{atom} + \epsilon_0 \hat x \cos(\omega_{IR} t)$$
where $T=2\pi/\omega_{IR}$ is the period of the classical electromagnetic field and $x$ is the polarization direction of the linear electromagnetic field. 
Within the acceleration gauge the Floquet Hamiltonian is given by,
$$\hat H_f\equiv -i\hbar \partial_t +\hat T+\hat V_{atom}({\bf r}+{\bf e}_x\alpha_0  \cos(\omega_{IR} t))$$
Where $\hat T= -\hbar^2/(2m_e)\nabla^2_{x,y,z}$ and $\alpha_0=\epsilon_0/({m_e\omega_{IR}})^2$. In our calculations for HHG, ATI and photon statistics we used the acceleration gauge approach.
The Floquet spectrum is obtained by solving the following eigenvalue problem
$\hat H_f|x,t\rangle_\alpha=E_\alpha|t\rangle_\alpha$ where
$|x,t\rangle_\alpha=\sum_{n_{f}=-\infty}^{+\infty} |\phi_{n_{f},\alpha}\rangle \exp(in_{f}\omega_{IR}t)$. To simplify the notation $\hat x$ includes the x-component of all the electronic coordinates  in the atom and $\phi_{n_{f},\alpha}$ is also a function of the y and z components of the positions of all electrons. In order to separate the Floquet resonances from the continuum the electronic coordinates are rotated into the complex plane by an angle $\theta$ (i.e., the electronic coordinates are scaled by a complex factor $\exp(i\theta)$). Upon complex scaling the Floquet spectrum become complex. The resonances Floquet solutions are those that are associated with  complex eigenvalues that are not affected by the value of $\theta$. Upon complex scaling the Flourier
components of the resonance Floquet solutions are square integrable function by using the c-prodcut \cite{NHQM-BOOK}, $\sum_{n_{f}=-\infty}^{+\infty}( \phi^\theta_{n_{f},\alpha}|\phi^\theta_{n_{f},\alpha})=1$ where $( \phi^\theta_{n_{f},\alpha}|\phi^\theta_{n_{f},\alpha})=\langle [\phi^\theta_{n_{f},\alpha}]^*|\phi^\theta_{n_{f},\alpha}\rangle$.
It is clear that every one of the $n-th$ Floquet channels is square integrable as well.  The complex scaled resonance Floquet solution that is mostly dominated by the field-free electron ground state of the atom is assigned  here as $\alpha=0$. The  width (i.e. inverse lifetime) of the resonance Floquet solutions is given by $\Gamma_{res}=-2Im E_{\alpha=0}$. To simplify the notation from now on  we will ignore the  $\alpha$ notation. That is, $|\phi^\theta_n\rangle$ stands for the Fourier component of the ground resonance Floquet state.  Within the framework of the length gauge approach the complex probability amplitude of transferring N infrared photons from the $n_f$ Floquet channel to the $n_f+N$ channel is given by\cite{NHQM-BOOK}, 
\begin{equation}
    {\cal{A}}(n_{f},N)=(n_{f}+N|\hat x|n_{f}) .
\end{equation}
However, we prefer to use the acceleration gauge here as it enables us not only to calculate the HGS but also the ATI spectra when the ionized electrons are not affected by the laser field. Within the framework of the acceleration gauge $ {\cal{A}}(n_{f},N)$ is given by,
\begin{equation}
    {\cal{A}}(n_{f},N)=(n_{f}+N|-\frac{dV(x)}{dx}|n_{f}) .
\label{PorbAmpNp}
\end{equation}
where $V(x)$ is an one electron potential which will be given later for our model of Xenon.
 The total complex  probability amplitude to absorb N infrared photons ( sometimes denoted here as $N_{IR}$) is
 \begin{equation}
     {\cal{A}}_{total}(N)=\sum_{n_{f}=-\infty}^{+\infty}  {\cal{A}}(n_{f},N) ,
 \end{equation}
 where $N$ gets odd value only due to the dynamical symmetry of the atom that interacts with electromagnetic waves (either classical or  quantum wave). The probability that the "loss" of  N infrared photons  will be detected by the infrared detector  is given by,
 \begin{equation}
 P(N_{IR})=|{\cal{A}}_{total}(N_{IR})|^2\equiv HGS(N_{IR})
 \end{equation}
 This result holds provided we neglect the population of several other Floquet excited resonance states that are also populated by the initial field free electronic ground state of the atom. Therefore, the HGS and the IR photon statistics  are exactly equal only within the single Floquet resonance approximation. In reality their relationship is more complicated (see Ref.\citenum{photon_statistics2017}). 
  The post-selection made by them in Ref.\citenum{photon_statistics2017} is on some diagonal in their diagram where the total energy of XUV and IR light is conserved.  This fits to our calculations of the HGS form a singe resonance Floquet solution that is associated with the energy of a mixed state of the atom and light. Therefore, our comparison between HGS and photon statistics is made from the calculations of a single resonance Floquet state. Due to the fact that the energy of the emitted XUV photons is given by $\hbar\omega_{IR}N_{IR}$, the HGS spectra is also given by $P(N_{IR})$.
  Therefore, the HGS can be obtained from the calculations (or measurements)   from   the number of absorbed IR photons, as measured by \textit{IR detectors}, which  is the same distribution of the number of XUV photons, as measured by \textit{XUV detector}. Provided, that  the same experiment is repeated many times while the number of absorbed infrared photons is a small fraction of the total number of the infrared photons emitted by the strong laser pulse with an envelop that supports  a large number of optical cycles. Notice that the ATI spectra are not involved in the photon statistics as the HGS, since the acceleration of the electrons is not involved in the ATI. Rather than emitting photons, electrons are ionized. See figure \ref{HGS-Xe} for the calculations of $P(N_{IR})=HGS$ for
   a model $Xe$ potential in
one spatial dimension with an atomic potential which has been
used before in the study of the high harmonic generation spectra in high-intensity laser fields\cite{fleischer2005adiabatic}. The Potential is given by,
\begin{equation}
    V(x)=V(x) = −0.63\exp(−0.1424x^2)
\end{equation}
This model atom supports a continuum, two bound states
with energies $E_1 = −0.44$ a.u., $E_2 = −0.14$ a.u, and a
third weakly bound state with energy $E_3 = −0.00014$ a.u.
To numerically represents the Floquet operator in matrix form
with a basis the time-periodic Fourier
basis functions, $\{\exp(i n_{f} \omega_{IR}t)\}_{n_{f}=0,\pm 1,\pm 2,..}$, and the field-free states in a particle in a box basis functions. $n_f$ stands for the index of the Floquet channel.
The laser parameters are $\omega_{IR}=0.0574$ a.u. and $\epsilon_0=0.015$ a.u.
\subsubsection{On the fluctuation in the number of absorbed infrared photons}
\begin{figure}[ht!]
   \centering
   \includegraphics[width=\columnwidth] {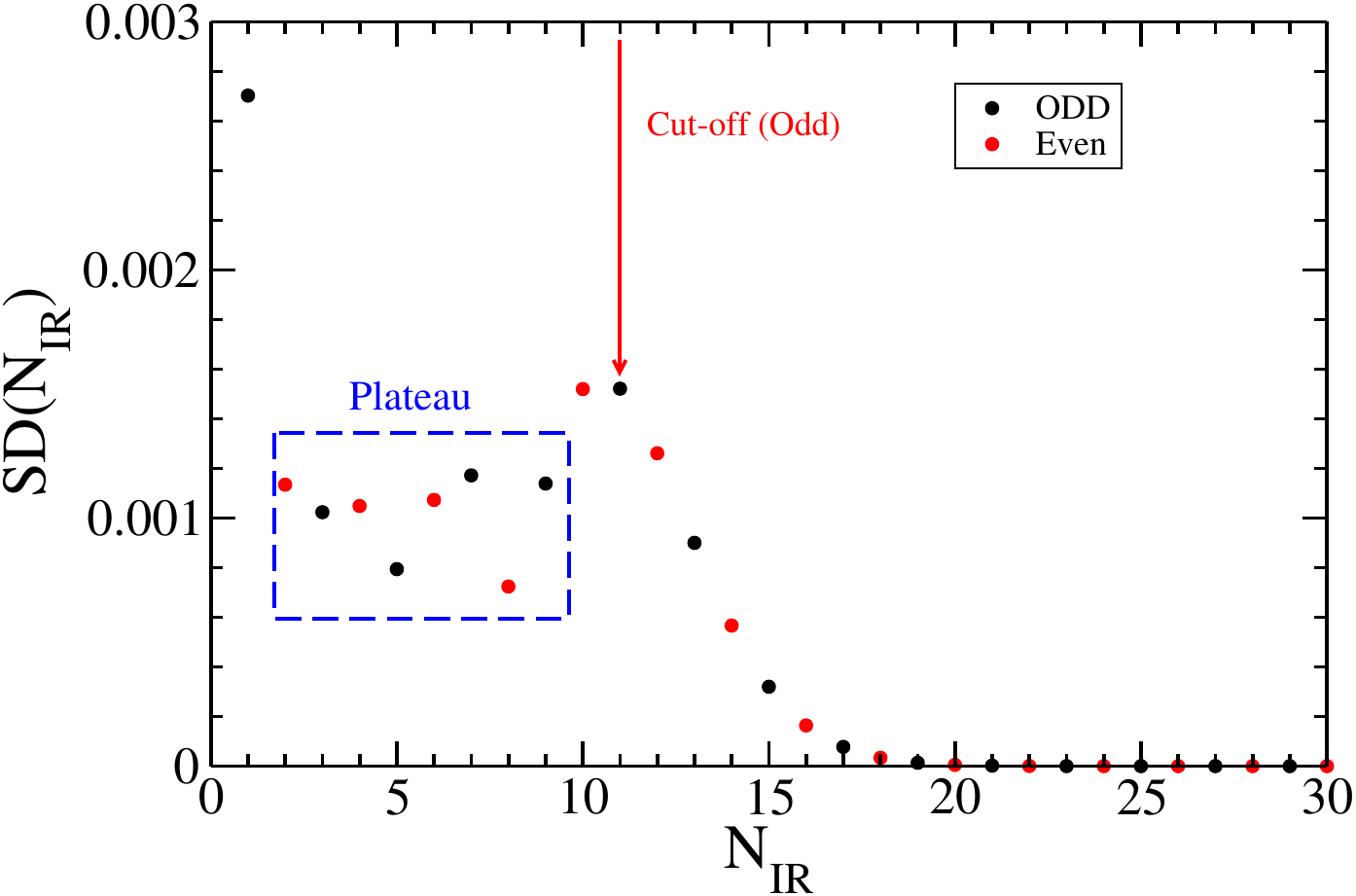}
    \caption{The  standard deviations in the number of absorbed IR photons,$N_{IR}$, as  a function of $N_{IR}$. Notice that the
      HGS is associated with the odd number of absorbed iR photons.
    }
   \label{SD-Xe}
    \end{figure}

As already discussed above the number of absorbed infrared photons that enables the emitted XUV radiation is odd.
The fluctuation in the number of IR  photons that  were absorbed  by the atom to generate XUV high order harmonics, is  given by the standard deviation (SD) expression 
\begin{eqnarray}
 && SD(N_{IR}) = \\ && \Bigg|\sqrt{\sum_{n_{f}=-\infty}^{+\infty}  \big[{\cal{A}}(n_{f},N_{IR})\big]^2-\big[\sum_{n_{f}=-\infty}^{+\infty}{\cal{A}}(n_{f},N_{IR})\big]^2}\Bigg|\nonumber
\end{eqnarray} 
where N is a odd number of infrared photons.\\
In Fig.\ref{SD-Xe} we see that the standard deviations of photon statistics for odd ones that are absorbed to emit the XUV radiation, can be divided into two groups. One group form a plateau with a cutoff at $N_{IR}=11$ (exactly as the cutoff in the HGS presented in Fig.{HGS-Xe}) and phase-lock photons with larger fluctuations (exactly as in the HGS). 


    
More information of the distribution of the infrared absorbed photons can be obtained by natural expansion of $ {\cal{A}} (n_{f},N_{IR})$\cite{moiseyev1986NATURAL_EXPANSION_analysis,moiseyev1986natural}. This issue is discussed below.  
\subsubsection{On the infrared photon distribution}
\begin{figure}[ht!]
     \centering
     \includegraphics[width=\columnwidth] {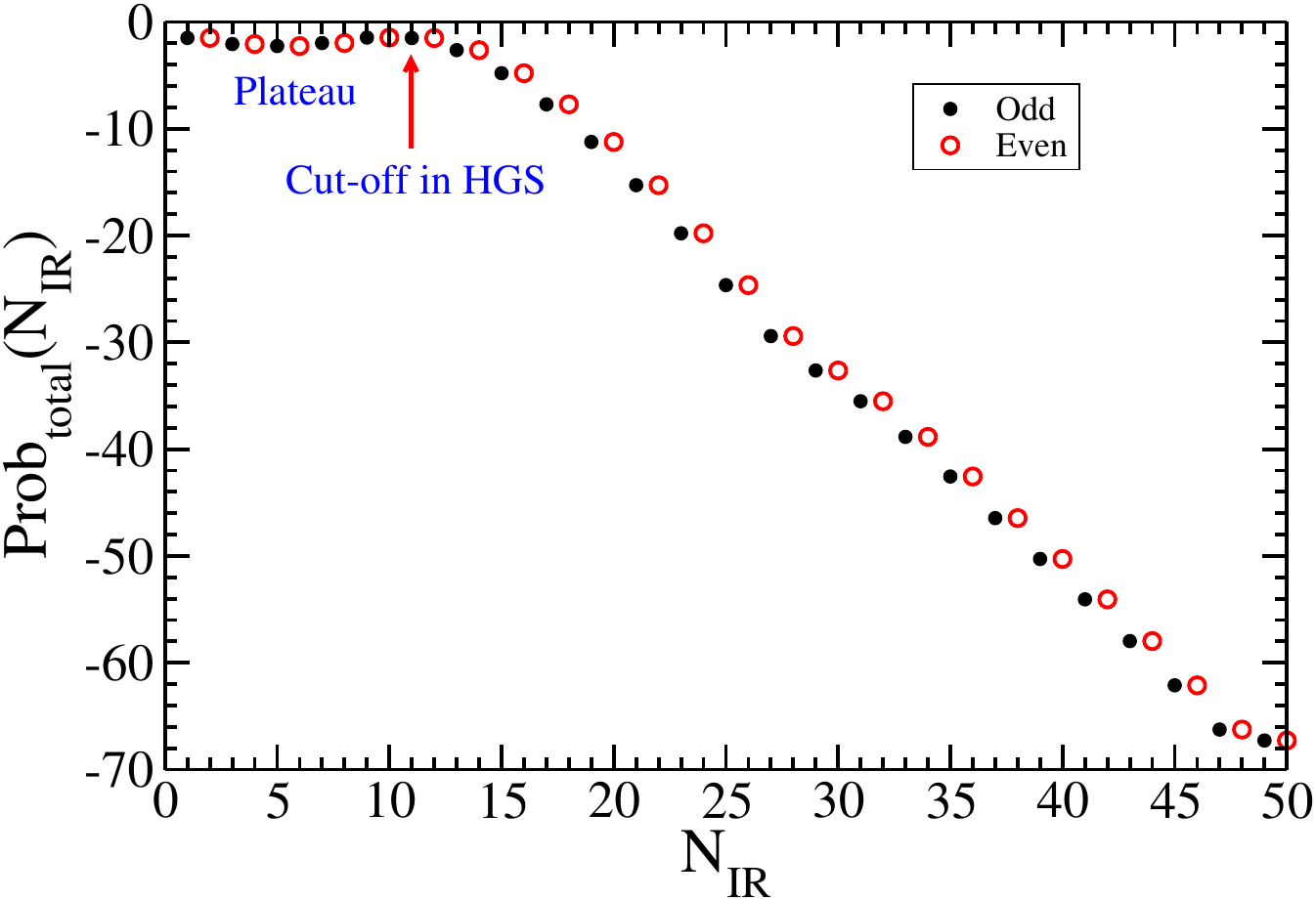}
      \caption{The fluctuations of absorbed IR photons for \textit{all} contributions to the natural expansion of ${\cal{A}}(n_{f},N_{IR})$.  See Eq.\ref{PROB_total} in the text.}
    \label{Fluctuation_total} 
    \end{figure}

The natural expansion\cite{moiseyev1986natural,moiseyev1986chaos-natural-expasnion} of the non-separable function
$ {\cal{A}}(n_{f},N_{IR}$ implies that,
\begin{equation}
{\cal{A}}(n_{f},N_{IR})=\sum_{l=1}^\infty d_l f_l(N_{IR})g_l(n_{f})
\end{equation}
where $|d_1|^2>|d_2|^2>...|d_l|^2...,n_f=0,\pm 1,\pm2,...$ are the Floquet channels and $N_{IR}=0,1,3,..$ are associated with the number of absorbed infrared photon. The expansion is ordered by the value of the occupation of each one of the terms in the expansion, ordered from largest to smallest of 
 $\big|d_l\big|^2=(Re[d_l])^2+(Im[d_l])^2$. The square of the orbital function $f_l(N_{IR})$ provides the the infrared photon distribution for the $l-th$ configuration. Therefore, the  \textit{partial}  infrared photon distribution is given by
 \begin{equation}
     Prob_l(N_{IR})\equiv |f_l(N_{IR})|^2 
     \label{PROB_partial}
 \end{equation}
where $l=1,2,...$ 
\begin{figure}[ht!]
     \centering
     \includegraphics[width=\columnwidth] {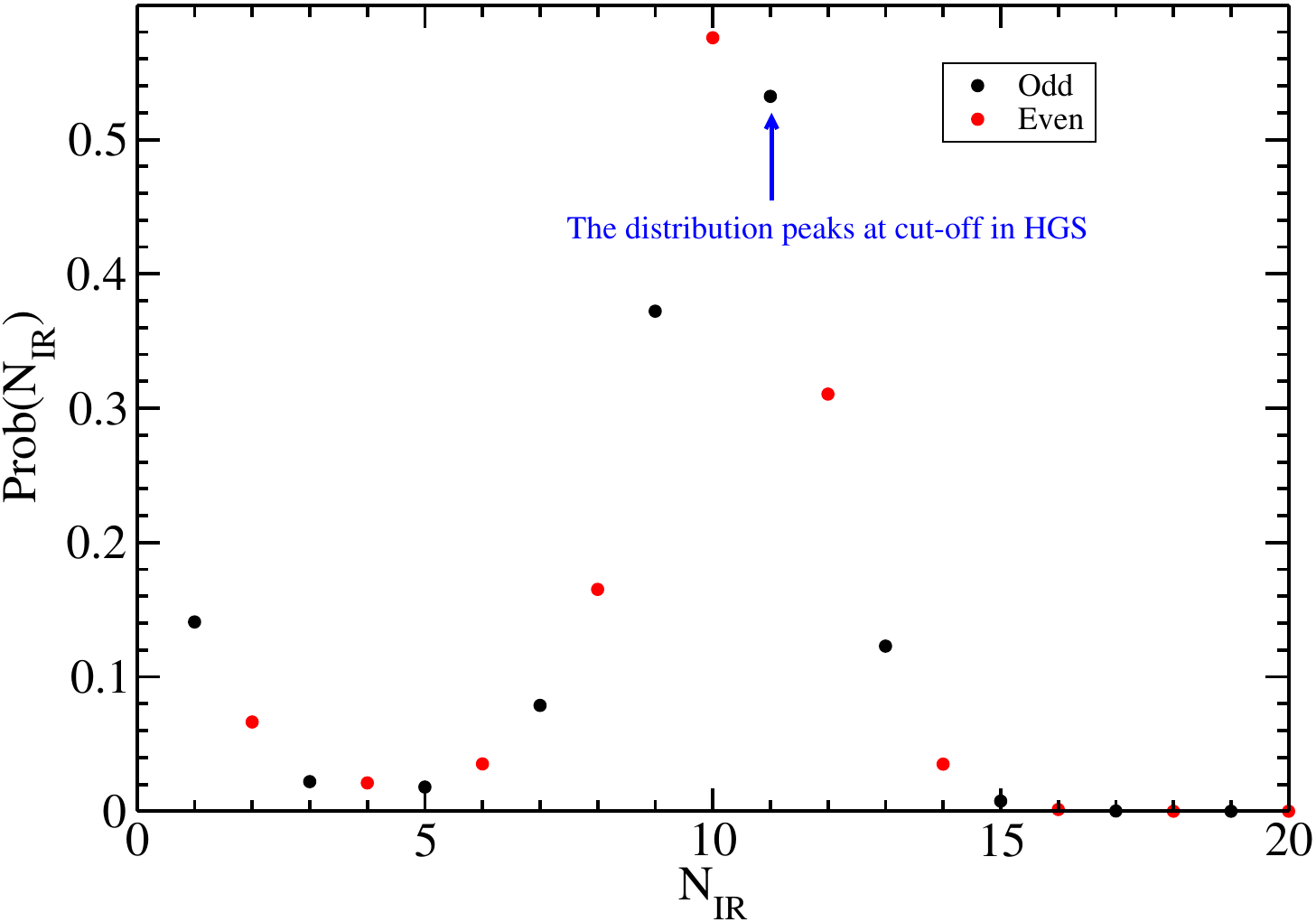}
      \caption{The  fluctuations of absorbed IR photons for the dominated contribution to the natural expansion of ${\cal{A}}(n_{f},N_{IR})$. See Eq.\ref{PROB_partial} in the text. Notice that this behavior is obtained only for the dominant term in natural expansion.}
    \label{Fluctuation_dominant}
    \end{figure} 
In Fig.\ref{Fluctuation_dominant} we show that the fluctuation in the photon statistics, as obtained from the leading term in the natural  expansion (i.e., $l=1$), is not Poisson as one might expect for classical light (see Ref.\citenum{photon_statistics2017}) but more like Wigner distribution which is a typical behavior for the finger print of chaos in spectroscopic measurements\cite{wigner1932quantum,kemp2012wigner}. See Ref.\citenum{averbukh1998CHAOS_HGS} on the association of the cutoff on the HGS with the classical chaotic dynamics.
The  \textit{total} infrared photon distribution is given by
 \begin{equation}
     Prob_{totoal}(N_{IR})\equiv \sum_{l=1}^\infty |d_lf_l(N_{IR})|^2 
     \label{PROB_total}
 \end{equation}
 The fluctuations presented in Fig.\ref{Fluctuation_total} is for the non-separable  function  $V={\cal{A}}(n_{f},N_{IR})$ that provides the complex probability amplitude for transfer  $N_{IR}$ red-photons from one Floquet channel to another. The plateau is as obtained in the HGS for negligible fluctuations in the amplitude of the measured high harmonics, whereas the fluctuations in the amplitudes of the high phase-lock harmonics which are observed beyond the cut-off are significantly larger.  These fluctuations result from the coherent nature of the Floquet solutions and would be observed even if one uses ideal detectors, in the absence of interactions among the atoms themselves and with the environment.

\subsubsection{ATI spectra and photon statistics}
\begin{figure}[ht!]
     \centering
     \includegraphics[width=\columnwidth] {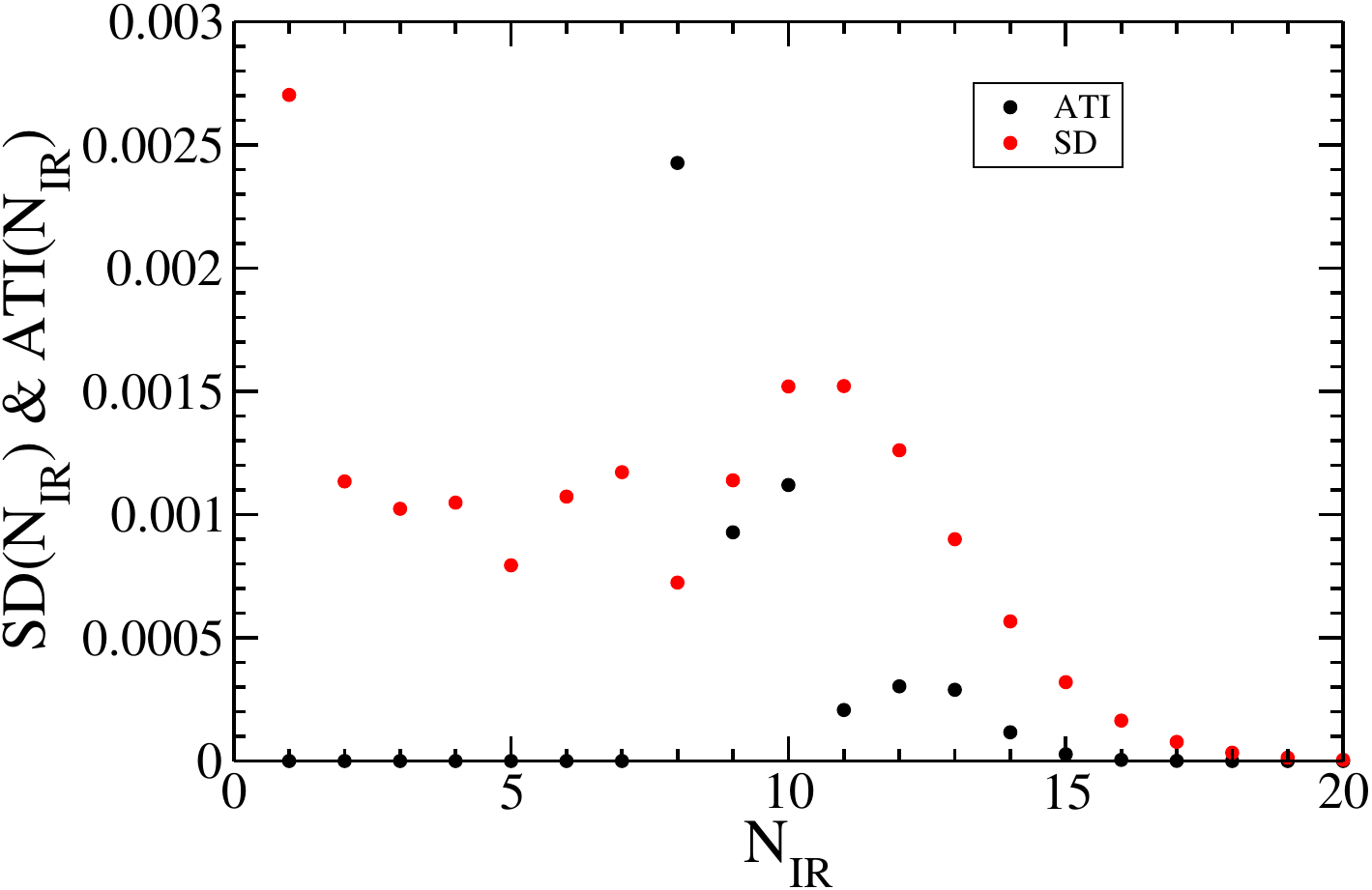}
      \caption{ATI and SD (standard deviation of IR photons) as a function of the number of absorbed infrared photons. The first peak in the ATI spectra is obtained for $N_{IR}=8$ which is the minimal number of absorbed  photons that enable ionization.}
    \label{ATI-SD}
    \end{figure}

The ATI spectra as the HGS and photon statistics are all calculated from a single resonance Floquet state that is dominated by the field-free ground state of the atom.
While in the HGS only odd number of IR photons are absorbed to emit  the XUV radiations the ATI spectra is obtained by abstained by absorbing odd and even number of photons.
Therefore, the competition between HGS and ATI happens only due the absorbing odd number of IR photons. Therefore, It is expected that the largest peak in the ATI spectra implies a minimal standard deviation of the absorbed photons since the ionization is fast (the height of the ATI peaks are related to the partial widths which are associated with the rate to ionization) and the lifetime of the photons that enable ionization is small. 
Indeed, Fig.\ref{ATI-SD} shows that the most pronounced  ATI peak for odd value of $N_{IR}$ is obtained for $N_{IR}=9$ where the standard deviation of the detected 9 absorbed photons gets a small value. The calculations of ATI spectra from a single resonance Floquet state are described in Ref.\citenum{peskin1994ATI} and in section 8.3.5 in Ref.\citenum{NHQM-BOOK}. 
The height of the peaks in the ATI spectra, within the acceleration gauge representation, are given by,
\begin{equation}
ATI({IR}) =\sum_{N_{IR}}\Gamma_{N_{IR}} ,
\end{equation}
where  $\Gamma_{N_{IR}}$ are the partial widths. 
The partial widths are given by,
$$\Gamma_{N_{IR}}=2(n_{floquet}=0+N_{IR}|-dV/dx|\psi_f)$$ where, 
$$(x|\psi_f)=\sqrt{\frac{m_e}{\hbar^2 k_{N_{IR}}}}e^{ik_{N_{IR}} x}$$ 
and $k_{N_{IR}}=\sqrt{2m_e(\hbar\omega_{IR}N_{IR}-|E_{bound}|)}>0$. The ground electronic energy of the atom is given by $E_{bound}<0$.
\section{Concluding remarks}
The post-selection conducted in Ref. \citenum{photon_statistics2017, gonoskov2016quantum} along a certain diagonal in their diagram appears to target the points where the total energy of XUV and IR light is conserved. This corresponds well with our calculations of High-Order Harmonic Generation (HGS), forming a single resonance Floquet solution associated with the energy of a hybrid state of the atom and light. Employing NHQM allows us to link both the HGS and the distribution of the infrared absorbed photons with single resonance Floquet solutions. Consequently, the HGS spectra, as detected by XUV detectors, can be obtained by monitoring the fluctuations of the infrared absorbed photons

\acknowledgments{ Mr. Matan Even-Tzur from the Hellen-Diller Quantum Center is acknowledged for bringing to my attention the open question regarding the correspondence between the high harmonic generation (HHG) measured by XUV spectrometer and by infrared (IR) spectroscopy measurements, and for his most helpful comments. Israel Science Foundation (ISF) grant No. 1757/24 is acknowledged for a partial support. }

\bibliography{References}

 \end{document}